\documentclass[aps,prx,onecolumn,showpacs,superscriptaddress]{revtex4-1}
\usepackage{amssymb}
\usepackage{amsfonts}
\usepackage{amsmath}
\usepackage{bm}
\usepackage{textcomp}
\usepackage{color}
\usepackage{longtable}
\usepackage{graphicx}
\usepackage{dcolumn}
\usepackage[a4paper=true,pagebackref=false]{hyperref}
\usepackage[normalem]{ulem}

\begin{document}

\title{Crystal field model simulations of magnetic response of pairs, triplets and quartets of Mn$^{3+}$ ions in GaN}

\author{D.~Sztenkiel} \email{sztenkiel@ifpan.edu.pl}
\affiliation{Institute of Physics, Polish Academy of Sciences, Aleja Lotnikow 32/46, PL-02668 Warsaw, Poland}

\author{K.~Gas}
\affiliation{Institute of Physics, Polish Academy of Sciences, Aleja Lotnikow 32/46, PL-02668 Warsaw, Poland}

\author{J.~Z.~Domagala}
\affiliation{Institute of Physics, Polish Academy of Sciences, Aleja Lotnikow 32/46, PL-02668 Warsaw, Poland}

\author{D.~Hommel}
\affiliation{Institute of Experimental Physics, University of Wrocław, M. Borna 9, Wrocław, Poland}
\affiliation{Lukasiewicz Research Network – PORT Polish Center for Technology Development, Stabłowicka 147, Wrocław, Poland}

\author{M.~Sawicki}
\affiliation{Institute of Physics, Polish Academy of Sciences, Aleja Lotnikow 32/46, PL-02668 Warsaw, Poland}

\date{\today}

\begin{abstract}

	A ferromagnetic coupling between localized Mn spins was predicted in a series of \textit{ab initio} and tight binding calculations and experimentally verified for the dilute magnetic semiconductor Ga$_{1-x}$Mn$_x$N. In the limit of small Mn concentrations, $x  \lesssim 0.01$, the paramagnetic properties of this material were successfully described using a single ion crystal field model approach. In order to obtain the description of magnetization in (Ga,Mn)N in the presence of interacting magnetic centers, we extend the previous model of a single substitutional Mn$^{3+}$ ion in GaN by considering pairs, triplets and quartets of Mn$^{3+}$ ions coupled by a ferromagnetic superexchange interaction.
Using this approach we investigate how the magnetic properties, particularly the magnitude of the uniaxial anisotropy field, change as the number of magnetic Mn$^{3+}$ ions in a given cluster increases from 1 to 4. Our simulations are then exploited in explaining experimental magnetic properties of Ga$_{1-x}$Mn$_x$N with $x \cong 0.03$, where the presence of small magnetic clusters gains in significance.
As a result the approximate lower and upper limits for the 
values of exchange couplings  between Mn$^{3+}$ ions in GaN, being in nearest neighbors $J_{\mathrm{nn}}$ and next nearest neighbors  $J_{\mathrm{nnn}}$ positions, respectively, are established.


$\newline$


\end{abstract}


\maketitle

\section{Introduction}

\baselineskip 15pt

	Magnetism in reduced dimensions, such as in magnetic nanostructures and magnetic clusters, has received a great research interest in the recent years due to its unexpected features and potential applications in high-density storage \cite{Sun:2000_Science}, nanoelectronics and quantum computations \cite{Troiani:2005_PRL}. A major advantage with respect to analogous bulk based materials originates from the additional degrees of freedom of nanoparticles to tune the magnetic properties by modifications of their size, shape, number of magnetic ions and/or coupling with the substrate \cite{Spisak:2002_PRB}.
For example, it was shown that single cobalt atoms deposited onto platinum (111) surface pose a very large magnetic anisotropy energy (MAE) of about 9~meV \cite{Gambardella:2003_Science}.	The single-ion MAE depends on the arrangements of atoms around the magnetic ion through the spin -- orbit interaction and the crystal field induced anisotropy of quantum orbital angular  momentum ($L$). In the bulk materials the magnitude of $L$ is usually quenched or strongly diminished by electron delocalization, ligand fields and hybridization effects, what result in small values of MAE of the order of 0.01~meV/atom \cite{Evans:2014_JPhysCM}. However, it is possible to enhance the MAE by using low-coordination geometries, such as atoms deposited on the surface, 1D atomic chains, magnetic clusters or molecular complexes. Values of MAE of the order of $1 \div 10$~meV per atom have been routinely reported in such systems \cite{Gambardella:2003_Science,Gambardella:2003_JPhysCM,Tung:2010_PRB}. Experiments on small particles of iron, cobalt, and nickel revealed a strong dependence of per-atom magnetic moments on the cluster size \cite{Billas:1994_Science}. The ferromagnetism was present even for clusters composed of about 30 atoms, with atom-like magnetization. The magnetic moments per one atom decreased with the number of ions in a given particle, approaching the bulk limit for about 500 atoms. Even in the material investigated here, the magnetic anisotropy strongly depends on the Mn ion concentration $x$, due to the dependence of the lattice parameters $c$ and $a$ of Ga$_{1-x}$Mn$_x$N on $x$. The magnetic anisotropy is high in a more diluted case \cite{Gosk:2005_PRB} and then decreases with $x$ \cite{Sztenkiel:2016_NatComm}. High MAE reduces the magnitude of the thermal fluctuations in superparamagnetic nanostructures and thus determines the potential applicability of these small-scale systems in high-density recording and magnetic memory operations. It is thus highly relevant to investigate the magnetic anisotropy properties of systems with reduced symmetry and/or coordination of magnetic aggregates.
	
		In this paper we numerically study how the MAE evolves from single isolated magnetic Mn$^{3+}$ impurity in GaN to very small magnetic clusters,  composed of up to four Mn$^{3+}$ ions coupled by ferromagnetic superexchange interaction. Here we take into account only the nearest neighbor (nn) interactions. Similar approach has been used to explain experimental results of single Mn ions and antiferromagnetically coupled pairs in InS-based dilute magnetic semiconductor (DMS) \cite{Tracy:2005_PRB}. In Ref \onlinecite{Soskic:2001_JMMM} the single ion crystal field model (CFM) was extended to simulate the magnetic properties of pairs and triplets, however the basis functions for diagonalization of the Hamiltonian were considerably restricted by taking only 10-fold degenerate functions of $^5$E symmetry. Here we use the CFM approach to model small magnetic clusters with up to 4 ions, where all function of $^5$E and  $^5$T symmetry are included in setting up the Hamiltonian with spin-orbit interaction and both trigonal and Jahn-Teller deformation taken into account.
	
	Importantly, the distribution of the different types of clusters in a random dilute magnetic semiconductor Ga$_{1-x}$Mn$_x$N can be precisely calculated for any value of $x$. This allows to compute the magnetization as a function of temperature $T$ or magnetic field $B$, and to compare it with the experimentally established one. Here we choose a Ga$_{1-x}$Mn$_x$N sample with $x \cong 3$\%, where the nn interactions play an important role but the long range percolation clusters (responsible of the ferromagnetism in this compound) are not yet statistically relevant.  As a result we give the approximate lower and upper limits for the values of exchange couplings  between Mn$^{3+}$ ions in GaN, being in nearest neighbors $J_{\mathrm{nn}}$ and next nearest neighbors  $J_{\mathrm{nnn}}$ positions, respectively.


\section{Model}

The standard theoretical approach to tackle magnetic systems is to use density functional theory (DFT) calculations that can give insight into the values of MAE, exchange integrals and atomic moments. Due to numerical complexity of the DFT, the simulations are generally limited to very small structures or bulk/2D periodic systems. In order to obtain macroscopic properties such the Curie temperature and total magnetization $M(T,B)$ of large systems one can resort to classical approximations, namely atomistic spin models supplemented with Monte Carlo or Landau-Lifshitz-Gilbert (LLG) dynamics \cite{Evans:2015_PRB}. However, small magnetic nanostructures at very low temperature are fundamental quantum mechanical systems due to the quantization of the relevant energy levels. For example, $M(T,B)$ characteristic of a single substitutional transition metal ion in a given semiconductor can be obtained using the full quantum-mechanical crystal field model approach. The CFM was developed by Vallin \cite{Vallin:1970_PRB,Vallin:1974_PRB} for II-VI dilute magnetic semiconductors doped with Cr, and then successfully applied for other DMSs \cite{Mac:1994,Twardowski:1993_JAP,Herbich:1998,Wolos:2004_PRB_b,Gosk:2005_PRB,Savoyant:2009_PRB,Stefanowicz:2010_PRB,Bonanni:2011_PRB,Rudowicz:2019_JMMM}. Recently it was shown that CFM simulations can explain the magnetic \cite{Gosk:2005_PRB,Stefanowicz:2010_PRB,Bonanni:2011_PRB}, magnetooptic \cite{Wolos:2004_PRB_b} and even magnetoelectric \cite{Sztenkiel:2016_NatComm} properties in dilute Ga$_{1-x}$Mn$_x$N, with  $x \leq 0.03$. Therefore, it is a natural way to extend aforementioned model of a single substitutional Mn$^{3+}$ ion in GaN by considering pairs, triplets and quartets of Mn$^{3+}$ ions coupled by ferromagnetic superexchange interaction \cite{Bonanni:2011_PRB, Sawicki:2012_PRB, Stefanowicz:2013_PRB}. Due to the fact that the number of elementary operations and computer memory needed for calculations grow exponentially with the number of particles, the CFM simulations are restricted here to magnetic clusters composed of up to four ions.

\begin{figure}[htb]
\centering
\includegraphics[width=7.6 cm]{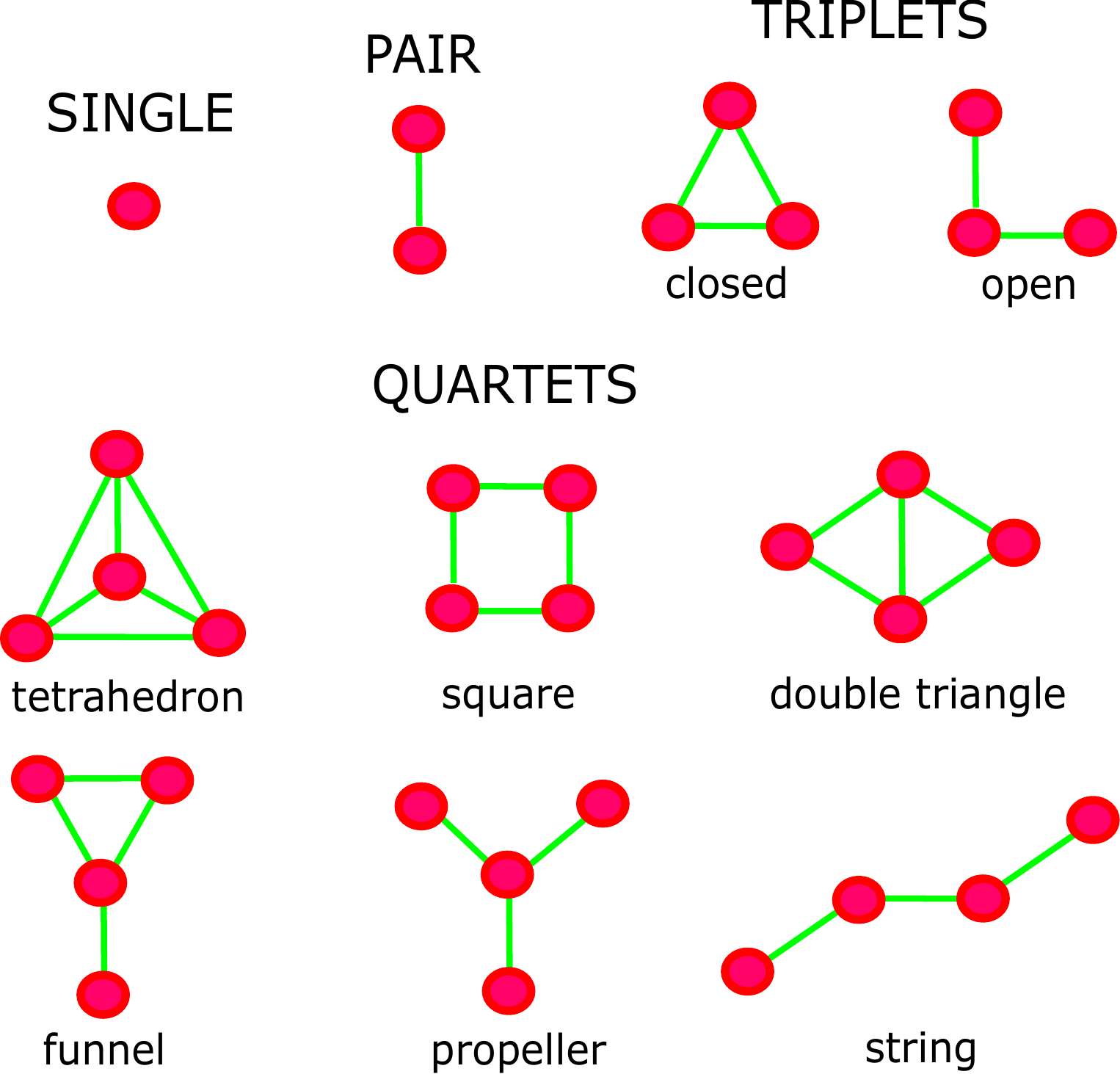}
\caption{\label{Fig:ClusterTypes} A magnetic cluster is defined here as a group of magnetic ions (red dots) coupled by
a nearest neighbor ferromagnetic superexchange interaction (green line), and decoupled from other more distant magnetic atoms. Only clusters with up to 4 ions are shown.}
\end{figure}

	The cluster types considered here are shown in figure~\ref{Fig:ClusterTypes}. Due to the short-ranged nature of the spin-spin interactions, only the couplings between nn Mn ions are taken into account. Each magnetic cluster is defined as a group of magnetic ions coupled by the nn ferromagnetic superexchange interaction, and decoupled from other more distant magnetic atoms. As Mn ions in Ga$_{1-x}$Mn$_x$N are \textsl{randomly} distributed over Ga cation sites \cite{Gas:2018_JALCOM}, the distribution of the different types of clusters, which depends on the Mn concentration $x$, can be precisely determined \cite{Shapira:2002_JAP}.

	The energy levels of a single Mn$^{3+}$ ion in wurtzite GaN are obtained by numerical diagonalization of the following (25 x 25) Hamiltonian matrix $H_S$ (see also Ref.~\onlinecite{Gosk:2005_PRB, Wolos:2004_PRB_b,Stefanowicz:2010_PRB,Sztenkiel:2016_NatComm})

\begin{eqnarray}
\label{eq:Hcf}
H_S(j)=H_{\mathrm{CF}}+H_{\mathrm{JT}}(j)+H_{\mathrm{TR}}+H_{\mathrm{SO}}+H_{\mathrm{B}},
\end{eqnarray}

where $H_{\mathrm{CF}}=-2/3B_4(\hat{O}_4^0-20\sqrt{2}\hat{O}_4^3)$
is the cubic field of tetrahedral $T_{d}$ symmetry, $H_{\mathrm{JT}}=\tilde{B}_2^0\hat{\Theta}_4^0+\tilde{B}_4^0\hat{\Theta}_4^2$
describes the static Jahn-Teller (J-T) distortion of the tetragonal symmetry, $H_{\mathrm{TR}}=B_2^0\hat{O}_4^0+B_4^0\hat{O}_4^2$ corresponds to the trigonal distortion along the GaN hexagonal $c$-axis, $H_{\mathrm{SO}}=\lambda\hat{\textbf{L}}\hat{\textbf{S}}$ represents the spin-orbit coupling. $H_B=\mu_{\mathrm{B}}(g_L\hat{\textbf{L}}+g_S\hat{\textbf{S}})\textbf{B}$ describes the Zeeman term where g-factors are $g_S=2$, $g_L=1$, $\mu_{\mathrm{B}}$ is the Bohr magneton and $B$ is the magnetic field. Here $\hat{\Theta}$ are Stevens equivalent operators for a tetragonal distortion along one of the three equivalent cubic $[100]$, $[010]$, $[001]$ directions denoted by $j=A, B, C$ respectively, and $\hat{O}$ are Stevens operators for a trigonal distortion along $[111]$~$\|$~$\textbf{c}$-axis of GaN. $B_i^k$, $\tilde{B}_i^k$, $\lambda_{TT}$, and $\lambda_{TE}$ denote parameters of the crystal field model, which are given in table~\ref{tab:Parameters_CF}. The numerical values of these parameters are taken from Ref.~\onlinecite{Stefanowicz:2010_PRB}.  

Interestingly, the CFM parameters can also be obtained from \textit{ab initio} supplemented with the superposition model calculations. In Ref.~\onlinecite{Virot:2010_JPCM} this procedure yielded the following values of the cubic field $B_4 = 11.67$~meV and the Jahn-Teller distortion parameters $\tilde{B}_2^0=-8.24$~meV and $\tilde{B}_4^0=-1.88$~meV. These values fit well with the CFM parameters presented in table~\ref{tab:Parameters_CF}.

\begin{figure}[htb]
\centering
\includegraphics[width=17cm]{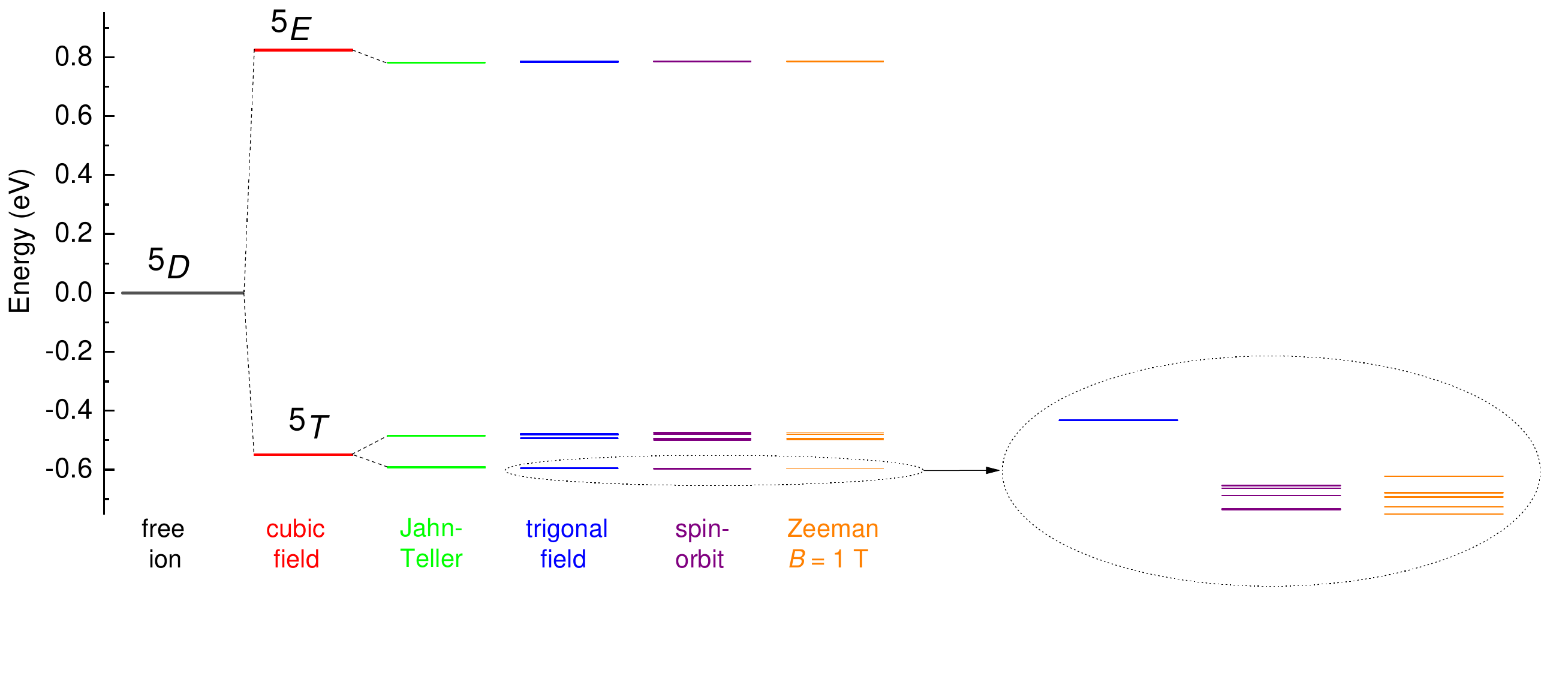}
\caption{\label{Fig:MultipletFinal} Energy level structure of Mn$^{3+}$ ion in GaN. The dashed lines represent the two first splittings by the cubic field and Jahn-Teller distortion. The magnetic field $B$ is applied along the $c$ axis of GaN.}
\end{figure}

The free Mn$^{3+}$ ion state is an orbital and spin quintet $^5$D with $L$=2 and $S$=2. The cubic field $H_{\mathrm{CF}}$ splits the $^5$D ground state into a tenfold orbital doublet $^5$E and 15-fold orbital triplet $^5$T, which is the ground state. The splittings of the $^5$D state of the Mn$^{3+}$ ion in GaN are shown in figure~\ref{Fig:MultipletFinal}. In general, the hybridization of the $d$ wave functions with the ligand wave functions is different for the $^5$E and $^5$T terms. Therefore, the three different parameters $\lambda_{TT}$, $\lambda_{TE}$ and $\lambda_{EE}=0$ (corresponding to the spin-orbit coupling between different combination of $^5$E and $^5$T states), are used in computations, instead of the single one $\lambda$ \cite{Herbich:1998, Gosk:2005_PRB}.

\begingroup

\begin{table}[htb]
  \centering \caption{Parameters of the crystal field model of the single Mn$^{3+}$ ion in GaN and superexchange interaction between nearest neighbors $J_{\mathrm{nn}}$ used in calculations. All values are in meV.}
\begin{tabular}{p{1.2 cm}p{1.2 cm}p{1.2 cm}p{1.2 cm}p{1.2 cm}p{1.2 cm}p{1.2 cm}p{1.2 cm}}

			\\
		
			\hline
            \hline
            	\\
      $B_4$ &  $B_2^0$   & $B_4^0$  &  $\tilde{B}_2^0$ & $\tilde{B}_4^0$ & $\lambda_{TT}$ & $\lambda_{TE}$ & $J_{\mathrm{nn}}$\\
      \\

      \hline

      \\

      11.44&1.1&-0.146&-5.1&-1.02&5.0&10.0&10.0\\
      \\
			\hline
			\hline

    \end{tabular}
  \label{tab:Parameters_CF}
\end{table}
\endgroup

	The basis of a single Mn$^{3+}$ ion (d$^4$ configuration, with $S=2$, $L=2$) consists of a total of W=$(2S+1)(2L+1)=25$ functions $|m_L,m_S\rangle$ characterized by spin $-2 \leq m_S \leq 2$ and orbital $-2 \leq m_L \leq 2$ quantum numbers. Due to the presence of three different J-T centers ($j=A, B$ or $C$), the average magnetic moment $\textbf{M}$ of Mn ion (in $\mu_{\mathrm{B}}$ units) can be calculated according to the formula
\begin{equation}
\label{eq:M_cf}
<\textbf{M}>=Z^{-1}\sum_{j=A,B,C}Z_j\textbf{M}^j,
\end{equation}
with $Z_j=\sum_{k=1}^W\mathrm{exp}(\frac{-E_k^j}{k_{\mathrm{B}}T})$  representing the partition function of the $j$-th center, $Z=Z_A+Z_B+Z_C$, and
\begin{equation}
\label{eq:M_cf_Center}
\textbf{M}^j=\frac{-\sum_{k=1}^W<\varphi_{k}^j|g_L\hat{\textbf{L}}+g_S\hat{\textbf{S}}|\varphi_{k}^j>\mathrm{exp}(-E_k^j/(k_{\mathrm{B}}T))}{Z_j},
\end{equation}
where $E_k^j$, $\varphi_{k}^j$ are the $k$-th eigenenergy and the  eigenfunction of the Mn$^{3+}$ ion being in $j$-th J-T center, respectively.

In this report we consider singles, pairs, triplets and quartets of Mn$^{3+}$ ions coupled by a ferromagnetic superexchange interaction $H_{exch}(1,2) = J_{\mathrm{nn}} \hat{\textbf{S}}_1 \hat{\textbf{S}}_2$.
The exact value of nearest neighbor superexchange coupling $J_{\mathrm{nn}}$ is not known.
The magnitudes of $J_{\mathrm{nn}}$ obtained from \textsl{first-principles} methods \cite{Sato:2010_RMP, Gonzales:2011_PRB} are rather high.  For example, according to \textit{ab initio} results, two nn ions are coupled by $J_{\mathrm{nn}} \cong 62$~meV \cite{Sato:2010_RMP} or $J_{\mathrm{nn}} \cong 55$~meV \cite{Gonzales:2011_PRB}. The same reports showed that the exchange coupling strongly depends on the Mn concentration and/or cluster type, as $J_{\mathrm{nn}}$  decreases to 15~meV for the case of tetrahedron quartet \cite{Gonzales:2011_PRB}. However, it is known that the \textit{first principles} results overestimate the values of the coupling between transition-metal ions \cite{Simserides:2014_EPJ}.
Recent Monte-Carlo simulations supplemented with tight binding parametrization, that described reasonably well the ferromagnetic properties of Ga$_{1-x}$Mn$_x$N with $1\% \leq x \leq 10\%$, used superexchange couplings up to 14-th neighbor with $J_{\mathrm{nn}}$=2.24~meV \cite{Simserides:2014_EPJ}.
Finally, as presented in the last section of this report, from the fitting of the results of our model to experimental data we estimate that $J_{\mathrm{nn}}\textbf{S}_1\textbf{S}_2 / k_{\mathrm{B}} \gtrsim 400$~K. Therefore we assume $J_{\mathrm{nn}}=10$~meV here. All values of superexchange coupling were given assuming that the Mn$^{3+}$ spins $\textbf{S}$ are classical vectors with norm $|\textbf{S}|$ = 2.

Now, the relevant eigenfunctions and eigenvalues are obtained by a numerical diagonalization of the full (25$\times$25), (25$^2$$\times$25$^2$), (25$^3$$\times$25$^3$), (25$^4$$\times$25$^4$) Hamiltonian matrix, for a single ion, pair, triplet or quartet, respectively. Additionally, one should take into account that the number of different J-T configurations increases with the number of ions $N$ in given cluster, and equals $3$, $3^2$, $3^3$ and $3^4$ for $N = 1, 2, ~3$, and 4, respectively. For example, the hamiltonian for an open triplet (c.f. figure~\ref{Fig:ClusterTypes}) reads
\begin{equation}
\label{eq:Hcf_closedtriplet}
H(j_1,j_2,j_3)=H_S(j_1,1) + H_S(j_2,2) + H_S(j_3,3) + H_{exch}(1,2) + H_{exch}(2,3),
\end{equation}
where $H_S(j,k)$ is the single ion hamiltonian for $k$-th ion being in the $j$-th J-T center and the base states are characterized by the set of quantum numbers $|m_{L_1},m_{S_1},m_{L_2},m_{S_2},m_{L_3},m_{S_3} \rangle$. Now, the magnetization of the cluster (in $\mu_{\mathrm{B}}$ units) is the thermodynamical and configurational average of the total magnetic moment operator $g_L(\hat{\textbf{L}}_1+\hat{\textbf{L}}_2+\hat{\textbf{L}}_3)+g_S(\hat{\textbf{S}}_1+\hat{\textbf{S}}_2+\hat{\textbf{S}}_3)$, and the sum $\sum_{j=A,B,C}$ in equation~(\ref{eq:M_cf}) is replaced by $\sum_{j_1=A,B,C}\sum_{j_2=A,B,C}\sum_{j_3=A,B,C}$.

	In order to speed up the calculations a parallelization of the code is used. Additionally, due to the presence of the Boltzman factor in equation~(\ref{eq:M_cf_Center}), only the lowest energy levels ($k_{max} < W$) and eigenfunctions need to be calculated in order to obtain a sufficiently good approximation of $M(B,T)$. Here $W$ is equal to $25$, $25^2$, $25^3$, $25^4$ for a single ion, pair, triplet or quartet, respectively.
At low temperatures, $T \lesssim 100$~K, the condition of $E_{k_{max}} - E_{k=0} > 22 k_{\mathrm{B}}$T is fulfilled, ensuring that relative error in calculation of $M(B,T)$ is practically zero. At the highest temperature $T = 400$~K, where $E_{k_{max}} - E_{k=0} > 6 k_{\mathrm{B}}T $ this error does not exceed $2\%$. $E_{k=0}$ is the ground state energy and $E_{k_{max}}$ is the maximal energy of the calculated excited states.

\section{Magnetic simulations}

\begin{figure}[htb]
\centering
\includegraphics[width=17 cm]{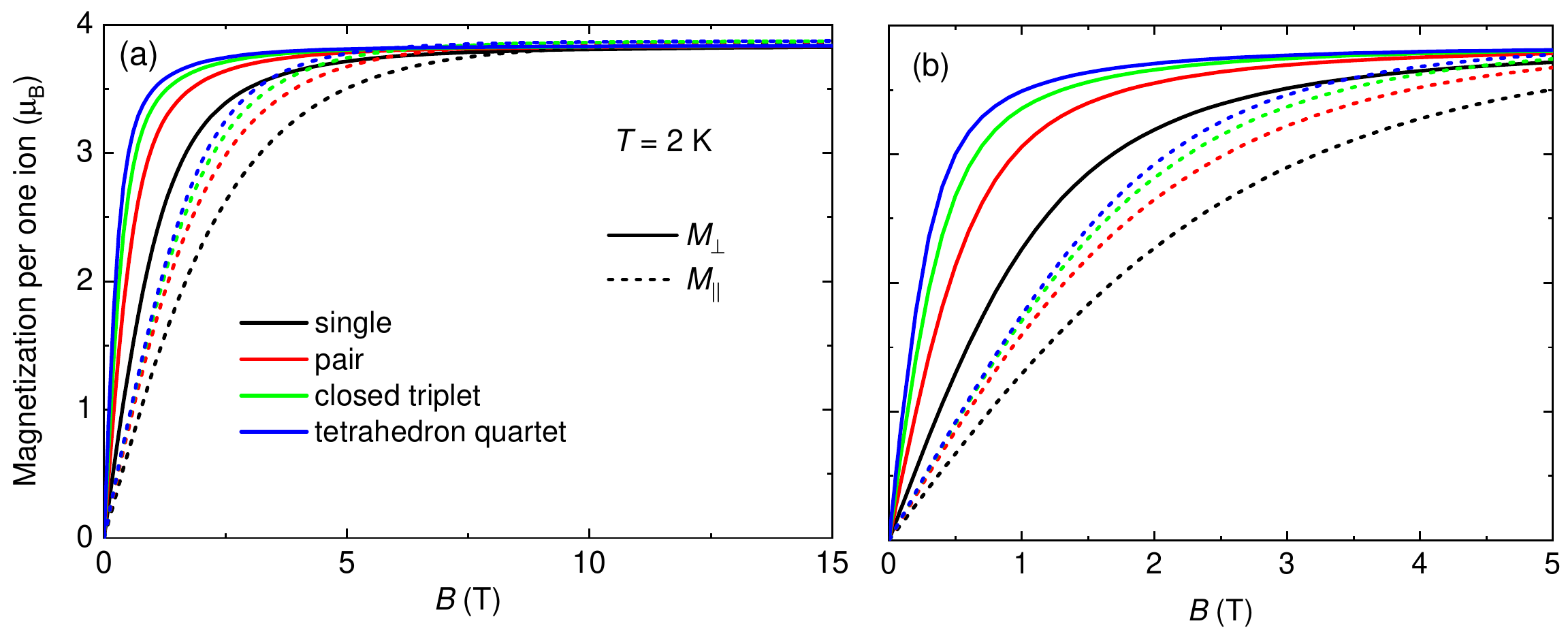}
\caption{\label{Fig:Magnetization} Magnetization per one ion as a function of the magnetic field $B$ of different Mn$^{3+}$ magnetic clusters in GaN at $T$=2~K obtained using crystal field model with ferromagnetic superexchange coupling $J_{\mathrm{nn}}=10$~meV. The magnetic easy axis $M_{\perp}$ (solid lines) is perpendicular to the $\textbf{c}$ axis of GaN, whereas the hard one $M_{||}$ (dashed lines) is parallel to the $\textbf{c}$ axis. (a) The high magnetic field, and  (b) the medium magnetic field region.}
\end{figure}

\begin{figure}[htb]
\centering
\includegraphics[width=9 cm]{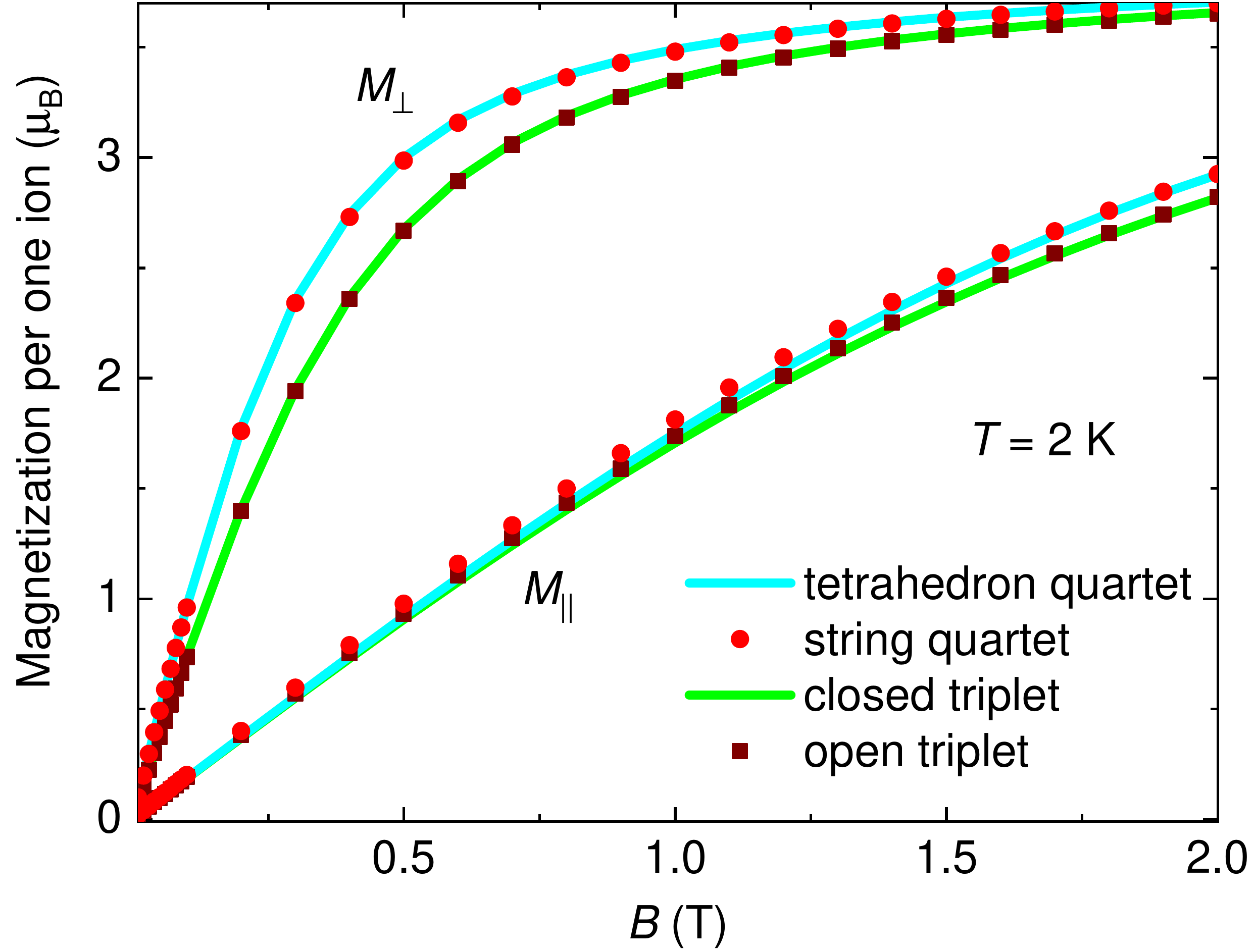}
\caption{\label{Fig:OpenVsClosed} Comparision of the magnetization per one ion of magnetic clusters composed of three ions (open vs closed triplet) and four ions (string vs tetrahedron quartet).}
\end{figure}

In figure~\ref{Fig:Magnetization} we present the results of our simulations of magnetization per one ion as a function of magnetic field of different magnetic clusters at $T=2$~K. We see that $M(B)$ varies sharply with magnetic field and the saturation is observed for high $B$. We choose such a low temperature, because a pronounced magnetic anisotropy is observed in (Ga,Mn)N only at $T \lesssim 10$~K. At a low to medium field range the magnetic easy axis $M_{\perp}$ is perpendicular to the $\textbf{c}$ axis of GaN ($\textbf{B}$ ${\perp}$ $\textbf{c}$), whereas the hard one $M_{||}$ is parallel to $\textbf{c}$ axis ($\textbf{B}$ $||$ $\textbf{c}$). Such uniaxial magnetic anisotropy is specific for Mn$^{3+}$ ions in wurtzite GaN \cite{Gosk:2005_PRB, Stefanowicz:2010_PRB, Sztenkiel:2016_NatComm}. As expected for ferromagnetic coupling between atoms, the magnetization per one ion increases with the size of the cluster $N$. Only one representative example from clusters with $N$ = 1, 2, 3, 4 is presented in figure~\ref{Fig:Magnetization}. As the strength of superexchange interaction is much stronger than the thermal energy $J_{\mathrm{nn}} \gg k_{\mathrm{B}}$T$ \approx 0.17$~meV at $T = 2$~K, the $M(B,T=2~K)$ curve of cluster with given $N$ is practically independent of the cluster type and the number of ferromagnetic bonds $J_{\mathrm{nn}}\textbf{S}_i\textbf{S}_j$. This is exemplified in figure~\ref{Fig:OpenVsClosed} where the same dependencies are observed for closed and open triplets as well as tetrahedron and string quartets. Therefore, in the remaining part of this report only the results corresponding to singles, pairs, closed triplets and tetrahedron quartets are considered.

 From figure~\ref{Fig:Magnetization} it can be seen that $M(B)$ clearly depends on the orientation of the magnetic field $B$ and the cluster size. On the other hand other studies do not show any conclusive dependencies of the magnetic anisotropy on the number of ions in given clusters or 1D wires. For example, experimental results on cobalt atoms deposited on an atomically ordered platinum surface \cite{Gambardella:2003_Science}, revealed that magnetic anisotropy decreases strongly with increasing Co coordination and the number of Co particles. The first-principles DFT investigations of nanometric Co$_n$Ni$_m$ clusters (with size $N = n + m \leq 7$) were performed in Ref.~\onlinecite{Mejia-Lopez:2018_PCCP}. A strong enhancement of MAE of the clusters as compared with bulk-like values was found. However, MAE of clusters as a function of their composition exhibited a complex and a non-monotonous behavior. These features were related to Co–Ni (spd) hybridization processes as well as structural rearrangements of the atoms.

\begin{figure}[htbh]
\centering
\includegraphics[width=17 cm]{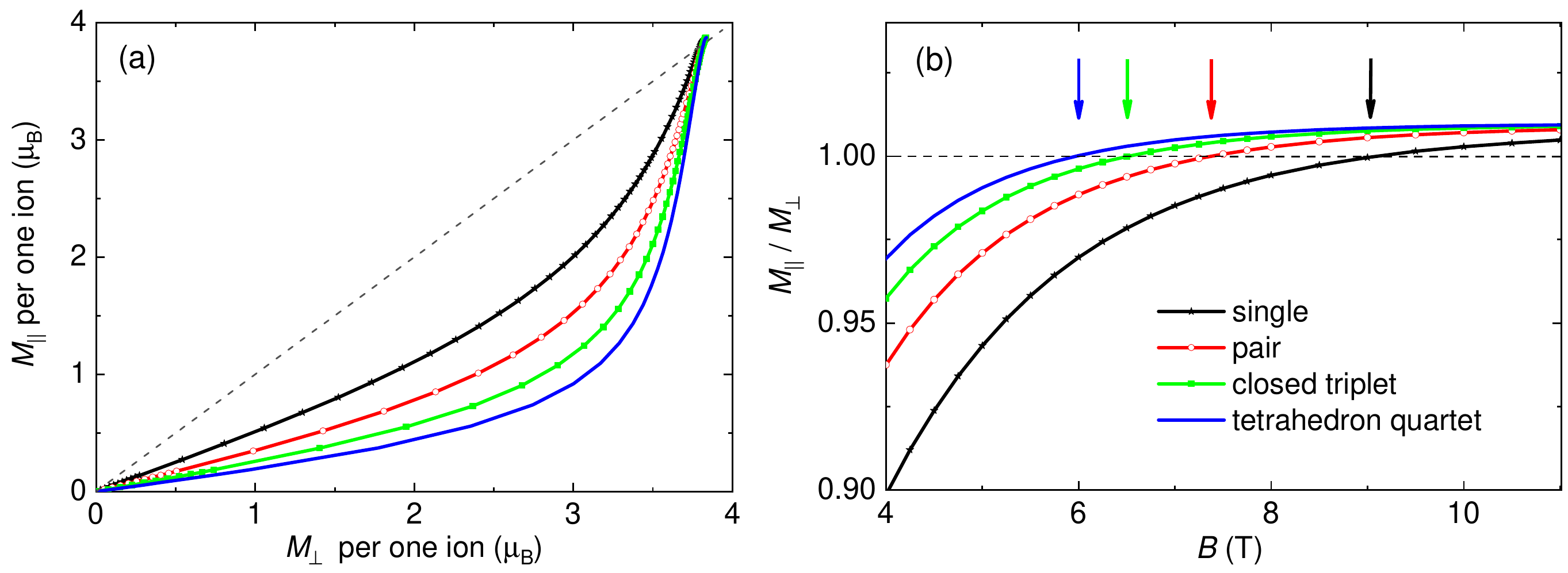}
\caption{\label{Fig:MzMx} (a) The hard axis magnetization $M_{||}$ as a function of the easy axis magnetization $M_{\perp}$. The isotropic case with $M_{||}$=$M_{\perp}$ is represented by the dashed line. (b) The $M_{||}/M_{\perp}$ ratio as a function of the magnetic field $B$. The uniaxial anisotropy fields, ${B}_{a}$ (marked by arrows), are established by the condition $M_{||}/M_{\perp}=1$ (dashed line).}
\end{figure}

A significant dependence of the magnetic anisotropy on the magnetic ion composition can also be observed in thin films. For example, the variations of MAE with concentration of Mn in (Ga,Mn)As grown on GaAs is usually caused by the epitaxial strain originating from lattice mismatch \cite{Fedorych:2002_PRB, Sawicki:2006_JMMM}.
A similar effect takes place in the material investigated here, where the deviation of the $c$-lattice parameter of the Ga$_{1-x}$Mn$_x$N from that of a GaN layer was observed as a function of the Mn content \cite{Kunert:2012_APL}. However, here, we have the advantage that all parameters of the single ion hamiltonian can remain unchanged during the transition from a singlet to quartets, and the only varying parameters are the number and the geometry of ferromagnetic bonds $J_{\mathrm{nn}}\textbf{S}_i\textbf{S}_j$.

\begingroup
\squeezetable

\begin{table}[htbp]
  \centering \caption{The uniaxial anisotropy field $B_a$ and magnetocrystalline energy (MAE) obtained from equation~(\ref{eq:MAE_integral}) for different Mn cluster types in GaN.} 
    \begin{tabular}{ p{2.5 cm}|p{2.0 cm}p{2.0 cm}p{2.0 cm}p{2.0 cm}}

			\hline
			\hline

       & single &  pair   & triplet  &  quartet\\

      \hline

      \\
			$B_a$ (T) &   9.2 &  7.3 &  6.5 &  6.0  \\

			\\
			
      MAE (meV) &  0.196 & 0.206 & 0.211 & 0.214  \\
			
			\hline
			\hline

    \end{tabular}
  \label{tab:MAE}
\end{table}
\endgroup


In order to quantify the strength of the magnetic anisotropy we use two different approaches. Firstly, we plot in figure~\ref{Fig:MzMx}(a) the hard axis magnetization $M_{||}$ as a function of the easy axis magnetization $M_{\perp}$. Obviously, the deviation of the $M_{||}=f(M_{\perp}$) curve from the isotropic case $M_{||}=M_{\perp}$ [the dashed line in figure~\ref{Fig:MzMx}(a)] indicates both the direction and the magnitude of the anisotropy. It is seen that the anisotropy increases with $N$, however this tendency is the most prominent between $N=1$ (single) and $N=2$ (pair).
Secondly, in the standard approach, we calculate MAE as the energy needed to rotate the magnetization from its easy axis into the hard one and it can be obtained from the following formula:
\begin{equation}
\label{eq:MAE_integral}
MAE=\int_0^{{B}_{a}}(M_{\perp}-M_{||})dB,
\end{equation}
where ${B}_{a}$ is the uniaxial anisotropy field [denoted by arrows in figure~\ref{Fig:MzMx} (b)]. The results of this procedure are summarized in table~\ref{tab:MAE}. A rather weak dependence of MAE on the cluster size is observed. Nevertheless, MAE increases clearly with $N$.
Interestingly, at the same time  ${B}_{a}$ decreases with $N$, as shown in figure~\ref{Fig:MzMx}~(b) and enumerated in table~\ref{tab:MAE}.
The second important result of these computations is that the CFM model predicts the reversal of the magnetic anisotropy at very high magnetic fields $B \gtrsim 6$~T.
The low-field easy axis is perpendicular to the $\textbf{c}$ axis ($M_{\perp}>M_{\parallel}$), but above $B_a$, instead of a saturation, $M_{\parallel}$ takes over $M_{\perp}$.
The existence of this spin reorientation transition on $B$ turns out to be  quite a robust effect in (Ga,Mn)N. Qualitatively the same behavior is obtained for different sets of CFM parameters, among others, for strongly enhanced uniaxial deformation parameters $B_2^0=3.2$~meV and $B_4^0=-0.42$~meV (not shown).

It is worth noting that in epitaxial Ga$_{1-x}$Mn$_x$N layers the magnetic anisotropy as a function of $x$ depends on two opposing effects.
Firstly, with increasing concentration of Mn$^{3+}$ ions, the number of small Mn clusters increases. This leads to the enhancement of MAE, as can be inferred from this study.
However, secondly, the lattice parameter $c$ of Ga$_{1-x}$Mn$_x$N epilayers increases with $x$ \cite{Kunert:2012_APL}.
The uniaxial magnetic anisotropy is controlled by the trigonal deformation characterized by parameter $\xi=c/a-\sqrt{8/3}$ \cite{Sztenkiel:2016_NatComm}, so with increasing $x$ the value of $\xi \rightarrow 0$, what reduces the magnitude of MAE.
In fact the latter is the dominant effect here.
As shown in the next section, a successful comparison of the CFM model to the experimental results for Ga$_{0.97}$Mn$_{0.03}$N requires reduced values of the uniaxial parameters $B_i^j$ from those used previously in more dilute material with $x < 1\%$ \cite{Stefanowicz:2010_PRB}.

\section{Comparison with experiment}

Previous studies of  Ga$_{1-x}$Mn$_x$N with $x \lesssim 1 \%$ revealed that for such a dilution the temperature-, field- and orientation-dependent magnetic properties
could have been adequately reproduced in the frame of the CFM approach limited solely to the case of non-interacting single Mn$^{3+}$ ions \cite{Stefanowicz:2010_PRB}.
However, with increasing $x$ the probabilities $p$ of Mn atoms to form nn pairs, triplets or quartets increases. For the random distribution case the magnitudes of relevant $p$ can be precisely calculated \cite{Shapira:2002_JAP}, so we can compare the results of the approach elaborated here with experimentally established magnetization $M_{\mathrm{Exp}}$ of (Ga,Mn)N of an adequately chosen Mn content. 
To this end we take $x \cong 3$\% sample for which  
nn interactions start to play an important role but the long range percolation clusters (responsible of the ferromagnetism in this compound) are not yet statistically relevant.
In order to model $M$ in this sample we take into account nn clusters up to the quartets.
We are limited in our approach by the exponentially increasing complexity of computation required for clusters with $N>4$.
The total magnetization computed in our method reads:
 \begin{equation}
\label{eq:M_aver}
M_{\mathrm{CFM}}=x N_0 ( p_1M_1+p_2M_2+p_{3o}M_{3o}+p_{3c}M_{3c}+p_{\geq4}M_{4t} ),
\end{equation}
where the magnitudes $M_i$: $M_1$, $M_2$, $M_{3o}$, $M_{3c}$ and $M_{4t}$ describe the magnetization per one ion of the single ion, pair, open triplet, closed tripled and tetrahedron quartet, respectively.
Each $M_i$ is calculated once for the relevant $B$- and/or $T$-range(s) and all are summed up with weights $p_i$ specific for the considered $x$.
The probabilities $p_i$ corresponding to $x = 3$\% are listed in table~\ref{tab:Probabilities}.
In our approach all clusters with number of ions $N \geq 4$ are represented by the tetrahedron quartet.
$N_0 = 4.4 \times 10^{22}$~cm$^{-3}$  is the cation concentration in GaN.

\begingroup
\squeezetable

\begin{table}[htbp]
  \centering \caption{ The probability $p$ that the Mn ion belongs to a given type of cluster. Only the nearest neighbor interactions count. The probability $p_{\geq4}$ represents the relative population of all clusters with a number of ions $N\geq4$.}
    \begin{tabular}{ p{2.0 cm}p{2.0 cm}p{2.0 cm}p{2.0 cm}p{4.0 cm}}
		
			\\		
			\hline
			\hline

           single &  pair   & open triplet  &   closed triplet  & $    \geq$ quartet\\
			\\
			
       $p_1$ &   $p_2$ &  $p_{3o}$ &   $p_{3c}$ & $p_{\geq4}$=1-$p_1$-$p_2$-$p_{3o}$-$p_{3c}$ \\
			
			\\

      \hline

      \\
			0.694 &  0.208 & 0.055 & 0.011 & 0.032  \\

			\hline
			\hline

    \end{tabular}
  \label{tab:Probabilities}
\end{table}
\endgroup

\begin{figure}[htbp]
\centering
\includegraphics[width=18 cm]{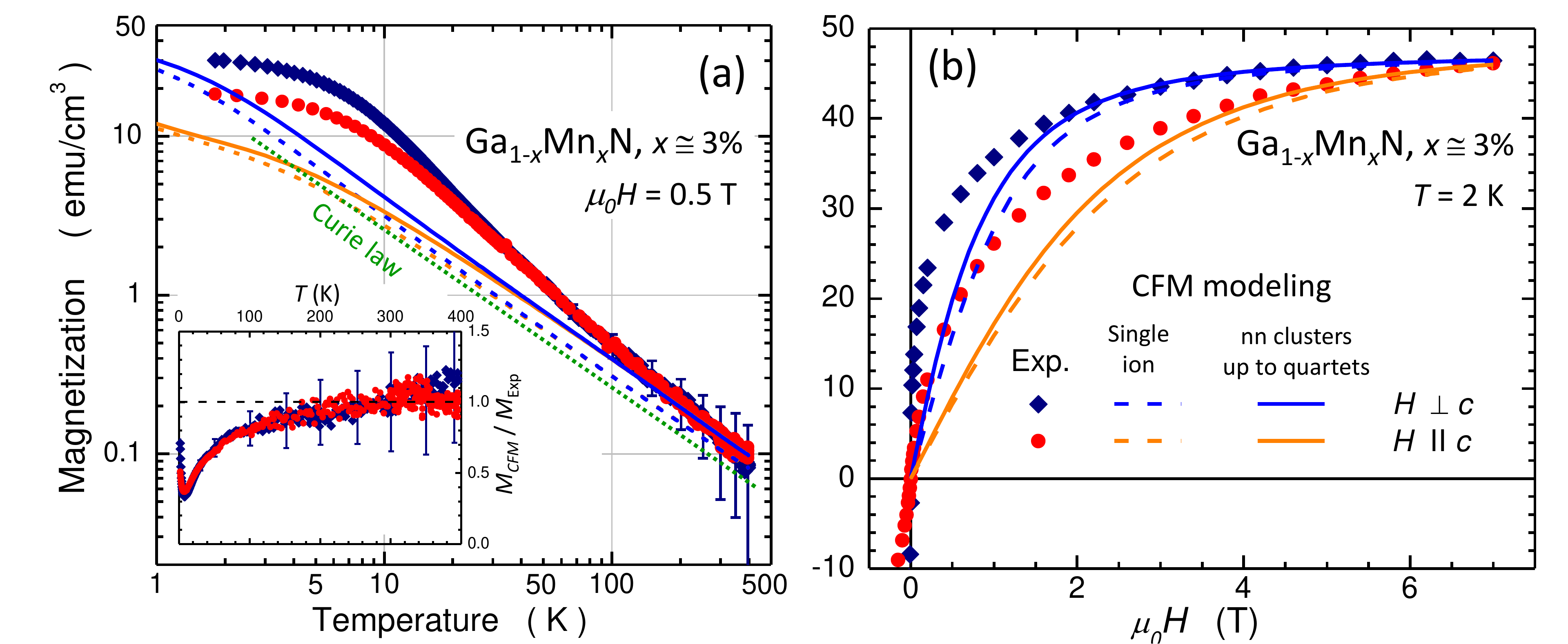}
\caption{\label{Fig:Exp} Comparison of the results of the crystal field model $M_{\mathrm{CFM}}$ and the experimentally established magnetization $M_{\mathrm{Exp}}$ in Ga$_{0.97}$Mn$_{0.03}$N as a function of temperature $T$, panel (a), and magnetic field $H$, panel (b). $M_{\mathrm{CFM}}$ results  for single uncoupled spins and nearest neighbor clusters (equation~\ref{eq:M_aver}) are denoted by dashed and solid lines, respectively. $M_{\mathrm{Exp}}$ is marked using symbols, defined in the legend in panel (b). The green dotted line represents the slope of the Curie law. The magnitude of the error bar indicated in the graphs is set primarily by the magnitudes of the standard deviation calculated for each point by the SQUID magnetometer software and the uncertainty of the ratio of the masses of the sample and of the reference piece of the substrate (about 1/500). In order to minimize the influence of other systematic errors (c.f. ref.~\onlinecite{Gas:2019_MST} for details) all four measurements required to establish the $T$-dependence of $M$ (two for each orientation for the sample and the reference) are performed immediately one after another at \textsl{the same} $H$, preset once for all four $T$-sweeps. On the other hand we disregard the uncertainty of the real volume of the (Ga,Mn)N layer. This systematic error is offset by such a choice of Mn concentration ($x \cong 3$\%) which allows $M_{\mathrm{CFM}}$ to reproduce the high-field magnetization data at low $T$ for $H \perp c$, panel (b). Accordingly, this procedure  yields the magnitude of $x$ in the layer.
Inset: The ratio of computed  magnetization $M_{\mathrm{CFM}}(T)$  to $M_{\mathrm{Exp}}(T)$. This ratio is calculated separately both for easy and hard magnetization direction. The input for the calculation is taken from the data plotted in panel (a). }
\end{figure}

The investigated here 190 nm thick Ga$_{0.97}$Mn$_{0.03}$N layer has been grown by plasma assisted molecular beam epitaxy (MBE) method in a Scienta-Omicron Pro-100 MBE according to previously elaborated protocols \cite{Gas:2018_JALCOM} on 3~$\mu$m thick GaN(0001) template layer deposited on a $c$-oriented sapphire substrate.
In order to avoid a high lateral gradient of $x$ caused by the inhomogeneous substrate temperature during the MBE growth \cite{Gas:2018_JALCOM}, the investigated specimen is cut form the center of the wafer where the Mn concentration and its homogeneity is the best \cite{Gas:2020_JALCOM}.
The magnetic measurements are performed using a Quantum Design MPMS XL Superconducting Quantum Interference Device (SQUID) magnetometer in the temperature range between 2 and 400 K and external magnetic field $H$, $|\mu_0 H | \leqslant 7$~T.
The measurements are carried out by strictly observing the experimental protocol for samples of minute signals deposited on bulky substrates \cite{Sawicki:2011_SST, Gas:2019_MST}.
The sample is measured in perpendicular ($H \parallel c$) and in-plane ($H \perp c$) orientation 
following the method described in \cite{Sztenkiel:2016_NatComm}.
The Mn concentration has been determined from the magnitude of the low--$T$ saturation magnetization (using the method outlined in the caption to figure~\ref{Fig:Exp}). In comparison of numerical results with experimental one we use $B$ and $H$ interchangeably, as for such dilute magnetic system the relative difference between these two values is within 3\%.

The parameters of the crystal field model (presented in table~\ref{tab:Parameters_CF}) are taken from Ref.~\onlinecite{Stefanowicz:2010_PRB}, where Ga$_{1-x}$Mn$_x$N with $x < 1 \%$ was studied, \textit{i.e.} the samples consisted predominantly of non-interacting (single) Mn$^{3+}$ ions. As described in Refs.~\onlinecite{Sztenkiel:2016_NatComm, Kunert:2012_APL}, the only one modification required here could be a change of the trigonal deformation parameters $B_i^j$ as they depend on wurtzite lattice parameters $c$ and $a$, which in turn depend on the $a$--parameter of the topmost layer of the substrate and on the Mn concentration in the layer \cite{Kunert:2012_APL}.
The parameters $B_i^j$ control the magnitude of the single ion uniaxial magnetic anisotropy in (Ga,Mn)N ~\cite{Sztenkiel:2016_NatComm}.
The relevant parameter here is $\xi=c/a-\sqrt{8/3}$, and, in the first approximation, $B_i^j \propto \xi$.
For the exercised here $x\cong 3$\% layer a high resolution x-ray diffractometry yields $\xi = -0.0027$. 
Accordingly, the best fit to experimental magnetization curves is obtained with $B_2^0 = 1.1$~meV and $B_4^0=-0.146$~meV. These values need to be smaller than those used previously to reproduce experimental $M$ and its anisotropy in very dilute Ga$_{1-x}$Mn$_x$N layers, which exhibit a much larger $\xi$ = -0.0077 \cite{Stefanowicz:2010_PRB}.

The quantitative comparison of the $T$-- and $H$--dependent magnetization $M_{\mathrm{Exp}}$ (symbols) and the computed $M_{\mathrm{CFM}}$ (lines of matching colors) is presented in figure~\ref{Fig:Exp}.
The two sets of lines indicate the single ion (no interaction) approach (dashed type) and nn clusters up to the quartets (solid type).
The $M_{\mathrm{Exp}}(T)$ taken at $\mu_0 H = 0.5$~T is shown in panel (a) in a double logarithmic scale.
This somewhat nonstandard way of presenting $M(T)$  brings out a $T^{-\alpha}$ dependence of $M$, a characteristic feature for random dilute paramagnetic compounds with a wide spectrum of exchange integrals \cite{Bhatt:1982_PRL}.
For random antiferromagnets $\alpha < 1$ \cite{Dietl:1987_JJAP, Sawicki:2013_PRB}, for random ferromagnets $\alpha > 1$ [e.g. (Ga,Mn)N)], and obviously $\alpha = 1$ for perfect paramagnets [exemplified in figure~\ref{Fig:Exp}~(a) by the dotted green line]. The experimental results for the $x \cong 3$\% sample indeed exhibit the $T^{-\alpha}$ dependence with $\alpha \cong 1$  for $T > 250$~K and with progressively increasing magnitude of $\alpha$ on lowering $T$ ($\alpha > 1$ between 20 and 250~K).
Such a behavior  clearly confirms the ferromagnetic nature of the superexchange coupling of Mn$^{3+}$ ions in (Ga,Mn)N.
We expand the discussion of the $T$--dependence of $M_{\mathrm{Exp}}$  at the last paragraph of this section.
Anyway, the results gathered in figure~\ref{Fig:Exp}~(a) clearly indicate that the single ion approach underestimates $M_{\mathrm{Exp}}(T)$ over the whole studied $T$--range, up to 400~K.
This fact strongly undelines the relevance of the nn coupling even above 300~K.
On the other hand, it is clearly seen, and substantiated in the inset, that the expansion of the CFM approach with nn clusters containing up to 4 ions results in an accurate reproduction of $M_{\mathrm{Exp}}(T)$ between 250 and 400~K.
A similar improvement in the reproduction of $M_{\mathrm{Exp}}(H)$ [exemplified for $T=2$~K in figure~\ref{Fig:Exp}~(b)] is seen in the mid-field region.

However, even this considerably improved approach sizably underestimates the $M_{\mathrm{Exp}}$ values at low temperatures and weak magnetic fields. This is a direct indication that both the much higher clusters and interactions between more distant neighbors are becoming more important at low $T$ and weak fields. 
Since the currently available computational resources are not able to cover the required expansion of the CFM model in any reasonable time frame, an another approach, namely the atomistic spin approach \cite{Evans:2014_JPhysCM,Evans:2015_PRB} seems to be the best option to properly reproduce $M_{\mathrm{Exp}}$.

On the other hand, the  comparison between $M_{\mathrm{CFM}}(T)$ and $M_{\mathrm{Exp}}(T)$ presented here provides a quantitative insight into the possible magnitudes of exchange couplings $J$.
According to figure~\ref{Fig:Exp}~(a), and in accordance with previous results for more diluted (Ga,Mn)N \cite{Stefanowicz:2013_PRB}, the slope of $M_{\mathrm{Exp}}(T)$ for $T < 250$~K deviates from the Curie law.
 The $1/T$ dependency  is expected to be obeyed by both the uncoupled spins $JS_1S_2 \ll k_{\mathrm{B}}T$ or by very strongly bound clusters $JS_1S_2 \gg k_{\mathrm{B}}T$.
 However, in the case when the exchange energy between more distant neighbors lies in the range of the experimentally accessible temperatures $2$~K $\leq T \leq 400$~K, that is when 2~K$ \leq  JS_1S_2 / k_{\mathrm{B}} \leq 400$~K, we observe an upturn of $M_{\mathrm{Exp}}(T)$ over the $T^{-1}$ line.
 As the temperature is decreased the clusters composed of next nearest neighbors, next next nearest neighbors and so on (decoupled at high $T$), start to significantly contribute to the total $M$.
 They are not included in our simulations.
 Contrary, at the highest temperatures, only the nn interaction is strong enough to overcome thermal fluctuations as the $M_{\mathrm{CFM}}(T)$ obtained from equation~\ref{eq:M_aver} describes very well the experimental data at $T \gtrsim 250$~K, as substantiated in the inset to figure~\ref{Fig:Exp}~(a).
 From these considerations we can approximate both the lower and the upper limits for  exchange couplings between nearest neighbors $J_{\mathrm{nn}}$ and next nearest neighbors ions $J_{\mathrm{nnn}}$, respectively. They read: $J_{\mathrm{nn}}S_1S_2 /  k_{\mathrm{B}} \gtrsim 400$~K and $J_{\mathrm{nnn}}S_1S_2 / k_{\mathrm{B}} \lesssim 250$~K.

\section*{Conclusions}
	
In this paper we numerically compute the magnetic response of small Mn$^{3+}$ magnetic clusters in GaN using a quantum-mechanical crystal field approach. The calculations are performed for isolated ions, pairs, triples and quarters of Mn$^{3+}$ ions coupled by nearest neighbor ferromagnetic superexchange interaction. 
We show that the magnetocrystalline anisotropy increases with the number of ions $N$ in a given cluster, whereas the uniaxial anisotropy field decreases with $N$.
Above this field the magnetic anisotropy is expected to change its sign assuming the direction of the wurtzite $c$ axis as the easy direction - an effect that awaits its experimental verification.
Our simulations have been also exploited in explaining experimental magnetic properties of Ga$_{1-x}$Mn$_x$N in the dilute case ($x \cong 0.03$), where different small magnetic clusters start to play an important role.
The direct comparison between the model computations and the experimental findings yields both the lower and the upper limits for  exchange couplings between nearest neighbors and next nearest neighbors ions, respectively.


\section*{Appendix}
\setcounter{equation}{0}
\renewcommand{\theequation}{A\arabic{equation}}

Here we calculate the Stevens operators entering the Jahn-Teller part of the Hamiltonian in the basis with a common $c$-axis. Assuming that the quantization axis is along $[111]$ $||$ $\textbf{c}$ direction ($\textbf{z}$ $||$ $[111]$), the Stevens operators read

\begin{equation}
\label{eq:O40a}
\hat{O}_{4}^0 = 3\hat{L}_z^2-L(L+1)
\end{equation}

\begin{equation}
\label{eq:O40b}
\hat{O}_{4}^2 = 35\hat{L}_z^4-30L(L+1)\hat{L}_z^2+25\hat{L}_z^2-6L(L+1)+3L^2(L+1)^2
\end{equation}

\begin{equation}
\label{eq:O40c}
\hat{O}_{4}^3 = 1/4 [ \hat{L}_z(\hat{L}_+^3+\hat{L}_-^3)+(\hat{L}_+^3+\hat{L}_-^3)\hat{L}_z ]
\end{equation}

On the other hand, Jahn-Teller distortion is along one of the three equivalent cubic $[100]$, $[010]$, $[001]$ directions denoted here by $\tilde{x}, \tilde{y}, \tilde{z}$ (in the main text by j = $A$, $B$, $C$) respectively. In cartesian coordinate system, with $\textbf{z}$ $||$ $\textbf{c}$, the three J-T vectors ($\tilde{x}, \tilde{y}, \tilde{z}$) are obtained from the $(x,y,z)$ ones in the following way

\begin{equation}
\label{eq:JTxx}
\tilde{e}_{x}=\sqrt{2/3}e_{x}+\sqrt{1/3}e_{z}
\end{equation}

\begin{equation}
\label{eq:JTxy}
\tilde{e}_{y}=-\sqrt{1/6}e_{x}-\sqrt{1/2}e_{y}+\sqrt{1/3}e_{z}
\end{equation}

\begin{equation}
\label{eq:JTxz}
\tilde{e}_{z}=-\sqrt{1/6}e_{x}+\sqrt{1/2}e_{y}+\sqrt{1/3}e_{z}
\end{equation}

For center $A$, assuming that J-T distrortion ${e}_{A}^{\mathrm{J-T}}$ $||$ $\tilde{e}_{x}$, the J-T momentum operator $\hat{\tilde{L}}_{z}$ and the terms $\hat{\tilde{L}}_{z}^2$, $\hat{\tilde{L}}_{z}^4$ appearing in $H_{\mathrm{JT}}$ now read

\begin{equation}
\label{eq:JTxx}
\hat{\tilde{L}}_{z}=\sqrt{2/3}\hat{L}_{x}+\sqrt{1/3}\hat{L}_{z}
\end{equation}

\begin{equation}
\label{eq:JTxz}
\hat{\tilde{L}}_{z}^2=2/3\hat{L}_{x}^2+1/3\hat{L}_{z}^2+\sqrt{2}/3(\hat{L}_{x}\hat{L}_{z}+\hat{L}_{z}\hat{L}_{x})
\end{equation}

\begin{multline}
\label{eq:JTxa}
\hat{\tilde{L}}_{z}^4=4/9\hat{L}_{x}^4+1/9\hat{L}_{z}^4 + 2/9[\hat{L}_{x}\hat{L}_{z}\hat{L}_{x}\hat{L}_{z}+\hat{L}_{z}\hat{L}_{x}\hat{L}_{z}\hat{L}_{x}+\hat{L}_{z}\hat{L}_{x}^2\hat{L}_{z}+\hat{L}_{x}\hat{L}_{z}^2\hat{L}_{x}+\hat{L}_{x}^2\hat{L}_{z}^2+\hat{L}_{z}^2\hat{L}_{x}^2] + \\ \sqrt{2}/9[\hat{L}_{z}^2(\hat{L}_{x}\hat{L}_{z}+\hat{L}_{z}\hat{L}_{x})+(\hat{L}_{x}\hat{L}_{z}+\hat{L}_{z}\hat{L}_{x})\hat{L}_{z}^2] + 2\sqrt{2}/9[\hat{L}_{x}^2(\hat{L}_{x}\hat{L}_{z}+\hat{L}_{z}\hat{L}_{x})+(\hat{L}_{x}\hat{L}_{z}+\hat{L}_{z}\hat{L}_{x})\hat{L}_{x}^2]
\end{multline}

Then Stevens operators for Jahn-Teller distortion along center $A$ are

\begin{equation}
\label{eq:Theta40a}
\hat{\Theta}_{4}^0 = 3\hat{\tilde{L}}_z^2-L(L+1)
\end{equation}

\begin{equation}
\label{eq:Theta40b}
\hat{\Theta}_{4}^2 = 35\hat{\tilde{L}}_z^4-30L(L+1)\hat{\tilde{L}}_z^2+25\hat{\tilde{L}}_z^2-6L(L+1)+3L^2(L+1)^2
\end{equation}

Similarly, one can calculate the expressions for Stevens operators for $B$ and $C$ types of centers if the chosen
quatization axis is along $c$ axis. That is, for center $B$ we have ${e}_{B}^{\mathrm{J-T}}$ $||$ $\tilde{e}_{y}$ and

\begin{equation}
\label{eq:JTxa}
\hat{\tilde{L}}_{z}=-\sqrt{1/6}\hat{L}_{x}-\sqrt{1/2}\hat{L}_{y}+\sqrt{1/3}\hat{L}_{z}
\end{equation}

and for center $C$ we have ${e}_{C}^{\mathrm{J-T}}$ $||$ $\tilde{e}_{z}$ with

\begin{equation}
\label{eq:JTxb}
\hat{\tilde{L}}_{z}=-\sqrt{1/6}\hat{L}_{x}+\sqrt{1/2}\hat{L}_{y}+\sqrt{1/3}\hat{L}_{z}
\end{equation}

Finally, simple (but lengthy) calculations will give us desired $\hat{\tilde{L}}_{z}^2$ and $\hat{\tilde{L}}_{z}^4$ terms.

\section*{Acknowledgments}

The work is supported by the National Science Centre (Poland) through project OPUS 2018/31/B/ST3/03438 and by the Interdisciplinary Centre for Mathematical and Computational Modelling at the University of Warsaw through the access to the computing facilities.






\end{document}